\documentclass[12pt,preprint]{aastex}
\shorttitle{Could  55 Cancri Really Be in the 3:1 MMR ? }
\shortauthors{Ji Jianghui et al.}

\begin{document}

\title{Could the 55 Cancri Planetary System Really Be in the 3:1 Mean Motion Resonance?}

\author{Jianghui JI\altaffilmark{1,3}, Hiroshi Kinoshita\altaffilmark{4}, Lin
LIU\altaffilmark{2,3}, Guangyu LI\altaffilmark{1,3}}
\email{jijh@pmo.ac.cn }

\altaffiltext{1}{Purple  Mountain  Observatory , Chinese  Academy
of  Sciences ,  Nanjing  210008,China} \altaffiltext{2}{Department
of Astronomy,  Nanjing University, Nanjing  210093, China}

\altaffiltext{3}{National Astronomical Observatory, Chinese
Academy of Sciences,Beijing 100012,China}
\altaffiltext{4}{National Astronomical Observatory,
 Mitaka, Tokyo 181-8588,Japan}

\begin{abstract}
We integrate the orbital solutions of the planets orbiting 55 Cnc.
In the simulations, we find that not only three resonant arguments
$\theta_{1}=\lambda_{1}-3\lambda_{2}+2\tilde\omega_{1}$,
$\theta_{2}=\lambda_{1}-3\lambda_{2}+2\tilde\omega_{2}$  and
$\theta_{3}=\lambda_{1}-3\lambda_{2}+(\tilde\omega_{1}+\tilde\omega_{2})$
librate respectively, but the relative apsidal longitudes
$\Delta\omega$ also librates about $250^{\circ}$ for millions of
years. The results imply the existence of the 3:1 resonance and
the apsidal resonance for the studied system. We emphasize that
the mean motion resonance and apsidal locking can act as two
important mechanisms of stabilizing the system. In addition, we
further investigate the secular dynamics of this system by
comparing the numerical results with those given by
Laplace-Lagrange secular theory.
\end{abstract}

%\keywords{celestial mechanics-methods:n-body simulations-planetary
%systems-stars:individual(55 Cnc, $\upsilon$ And, GJ 876, HD 82943,
%47 Uma, HD 12661) }

\section{Introduction}
To date, 87 planetary systems, containing 101 giant extrasolar
planets, have been discovered in Doppler surveys of solar-type
stars (Butler et al. 2003), among which there are 10
multiple-planet systems, including 8 two-planet systems (HD 82943,
GJ 876, HD 168443, HD 74156, 47 Uma, HD 37124, HD 38529 and HD
12661) and 2 three-planet systems (55 Cnc and $\upsilon$ And). As
a second newly-discovered triple planet system, the most
interesting characteristic of 55 Cnc is that the ratio of the
orbital periods of the inner two companions (Marcy et al. 2002) is
$44.27/14.65\simeq 3$, which suggests they are near  3:1 Mean
Motion Resonance (MMR). However, the 3-Keplerian fits and other
results given by Marcy et al. (2002) do not imply any evidence of
the resonance for 55 Cnc at present. Here we examine revised
orbital elements from Fischer et al.(2003) to search for evidence
of a 3:1 resonance. Next, if the two planets are in the resonance,
what kinds of the mechanisms make the system stable? In the
present work, we aim to investigate the problem.

We completed pure N-body codes (Ji, Li \& Liu 2002) to perform
numerical simulations for 55 Cnc system by using RKF7(8)(Fehlberg
1968)and symplectic integrators(Feng 1986;Wisdom \& Holman 1991).
The time step is 0.36 d, about $\sim$2.5 percent of the orbital
period of the innermost planet,which is sufficiently small for the
integration. In addition, the numerical errors were effectively
controlled with the accuracy  of $10^{-14}$ over the timescale
from 1 Myr to several million years.

\section{Dynamical Analysis}
Two orbital solutions for the 55 Cnc have appeared in the
literature: the first data set was reported by Marcy et al.
(2002), and with more observations, Fischer et al. (2003) further
updated the orbital parameters of 55 Cnc system. Their
publications allow us to obtain the values for the integration. In
this paper, we adopt the orbital parameters of the system provided
by Fischer et al. (2003). In the simulation, the mass of the
parent star is $1.03M_{\odot}$, and those of three orbiting
planets are adopted as $0.83M_{jup}$, $0.20M_{jup}$ and
$3.69M_{jup}$ respectively (see Table 1), by assuming $\sin i=1$.
In preparation of the orbital elements for integration, we assume
that all of the planets are initially in the same plane, then fix
their semi-major axes and eccentricities and the remainder angles
(nodal longitude, argument of periastron, mean anomaly) are
randomly generated for each orbit. In total, we carried out
hundreds of the simulations for various initial conditions for the
55 Cnc system. More than 90 percents of the systems that are not
linked to the 3:1 MMR become unstable in thousands of years. In
Table 1, we list one set of the values out of 8 orbital solutions
in association with the 3:1 MMR and apsidal resonance.

To examine our orbital solution based on Fischer et al. (2003), we
used our orbits to compute the radial velocity. In Fig.1, the
upper panel shows the comparison of the measured radial velocities
and theoretically calculated radial velocities for 55 Cnc-the sign
diamond represents the observations given by Marcy et al.
(2002)(see their Table 2), and the sign plus denotes the results
from our orbital solution; next we look at the lower panel which
exhibits the residual velocity with rms of 11.50  m s$^{-1}$.

\subsection{3:1 Mean Motion Resonance}
Many authors have studied the mean motion resonance in the
multiple-planet systems: GJ 876 (Laughlin \& Chambers 2001;
Kinoshita \& Nakai 2001; Lee \& Peale 2002; Ji, Li \& Liu 2002)
and HD 82943 (Gozdziewski \& Maciejewski 2001) are respectively in
the 2:1 MMR, 47 Uma (Laughlin, Chambers \& Fischer 2002) is close
to a 7:3 commensurability. And we are now concentrating on the
dynamics of the 3:1 resonance of 55 Cnc. The critical arguments
for the 3:1 MMR are
$\theta_{1}=\lambda_{1}-3\lambda_{2}+2\tilde\omega_{1}$,
$\theta_{2}=\lambda_{1}-3\lambda_{2}+2\tilde\omega_{2}$ and
$\theta_{3}=\lambda_{1}-3\lambda_{2}+(\tilde\omega_{1}+\tilde\omega_{2})$,
where $\lambda_{1} $, $\lambda_{2} $ are the mean longitudes of
Companions B and C respectively, and $\tilde {\omega}_{1}$,
$\tilde {\omega}_{2}$ respectively denote their periastron
longitudes (hereafter subscript 1 denotes Companion B; 2,
Companion C).

Marcy et al. (2002) pointed out that they did not find the
evidence of the 3:1 MMR by showing the time behavior of
$\theta_{1}$ with triple-Keplerian fit and self-consistent fits,
for no libration of the resonant argument was detected. They also
argue that further observations and dynamical fitting will clarify
the relationship between the two inner planets of 55 Cnc. By
contrast, we performed the integration on the basis of the fitting
by Fischer et al. (2003). Figures 2a--2c exhibit the arguments of
$\theta_{1}$, $\theta_{2}$ and $\theta_{3} $ plotted against time.
From the figures, we see that $\theta_{1}$ librates about
$180^{\circ}$, $\theta_{3}$ walks around $0^{\circ}$ with large
amplitudes, and $\theta_{2}$ wanders near $90^{\circ}$ with an
amplitude of $\pm70^{\circ}$ for the timescale of 3 Myr. Because
of the existence of the 3:1 resonance, as seen in Fig.3, the upper
panel shows the semi-major axis of Companion B $a_{1}$ executes
small vibrations about 0.115 AU, while the lower panel displays
that of Companion C $a_{2}$ varies about 0.240 AU for millions of
years. Obviously, these results confirm that the two companions of
55 Cnc are now really in the 3:1 MMR. Here we simply present the
fact of the resonance in the sense of the numerical results, our
future work will make an analytical theory to explain the orbital
resonance for this planetary system.

However, one may ask what the situation was like before the two
companions enter into the resonance. At first, we recall a recent
work on the likely origin of GJ 876 by Lee \& Peale (2002). They
revealed the differential planet migration due to planet-nebular
interaction could lead to capture into the 2:1 MMR. If we do not
consider the outermost Companion D with a long period of 4780
days, the 55 Cnc system is quite similar to GJ 876, such as the
semi-major axes and the masses of resonant pairs. In fact, the
real system is not the case, because of the presence of Companion
D, the process of the migration would possibly become more
complicated. Future studies (Peale, in private communication) will
help unearth the enigma of the origin of the resonance for 55 Cnc.

\subsection{Secular Dynamics of 55 Cnc}
For the system with small eccentricities and inclinations, the
long-term variations of the eccentricities (or inclinations) of
the planets can be described by the classical
Laplace-Lagrange(L-L) secular theory provided that the mean motion
resonance is absent. Now we draw attention to the secular behavior
of 55 Cnc system. In Table 2 are given the data from the secular
theory by considering all three planets. We see the contributions
from Companion D is much less than one percent of the mutual
effects of two inner planets. Hence, we can discuss the secular
dynamics of Companions B and C by ignoring Companion D, without
changing the quantitative results.

The linear L-L secular solutions for the eccentricities(Murray \&
Dermott 1999)can be written:
\begin{equation}
\label{eq1}    {
\begin{array}{l}
e_{j}\sin\tilde\omega_{j} = e_{j1}\sin(g_{1}t+\beta_{1}) + e_{j2}\sin(g_{2}t+\beta_{2}),\\
e_{j}\cos\tilde\omega_{j} = e_{j1}\cos(g_{1}t+\beta_{1}) + e_{j2}\cos(g_{2}t+\beta_{2}),\\
\end{array}}
\end{equation}
where $e_{ji}$, $g_{i}$ and $\beta_{i}$ ($j,i=1,2$, $j=1$ for
Companion B, $j=2$ for Companion C) can be evaluated by the
initial conditions in Table 1, and $t$ denotes time. For this
case, we obtain two eigenfrequencies $g_{1}=0.727^{\circ}$/yr and
$g_{2}=0.128^{\circ}$/yr. Together with $\beta_{1}=-4.92^{\circ}$,
$\beta_{2}=-23.46^{\circ}$, and other determined constants in
equation (\ref{eq1}), we can derive the analytical expressions of
the eccentricities $e_{j}$ for both of the inner planets. The
numerical eccentricities of Companions B and C are  respectively
in the intervals of [0.0, 0.22] and [0.20, 0.41], while the
theoretical values belong to those of [0.0, 0.17] and [0.30,
0.41],respectively. Because the inner two planets are now trapped
in the 3:1 resonance, the secular theory is expected to be
departure from the numerical results (we will construct a more
detailed model under the consideration of the 3:1 MMR to study the
system in our forthcoming paper), but it does correctly
demonstrate the qualitative long-term variations. Therefore, we
still perform integrations for the system that only consists of
two inner planets to numerically examine the long-term orbital
evolution without Companion D. We again find the outermost planet
will not greatly influence on the secular motion of the inner
companions because Companion D is fairly distant from Companions B
and C(the ratio of $a_{2}/a_{3}\simeq0.04$, $a_{2},
$$a_{3}$ are the semi-major axes of Companions C, D) and the
mutual perturbations are very small. Furthermore, with additional,
longer integrations of three planets, we find that the
eccentricity of Companion D ranges from 0.26 to 0.29 and its
secular motion is also steady. No sign shows this system is
chaotic by numerical results as well as secular theory. Hence, we
can safely conclude that the 55 Cnc system is dynamically stable
for the lifetime of the star.

To explore the stability of 55 Cnc with respect to the planet
masses, we fix $\sin i$ in increment of 0.1 from 0.1 to 0.9. In
the complementary experiments, we integrate the system for 1 Myr
with rescaled masses and the orbital parameters in Table 1. As a
result, we find the system become unstable in short time when
$\sin i<0.5$, indicating more massive planets are not fit for this
system.

\subsection{Apsidal  Resonance}
Apsidal resonance occurs when two planets share the same
time-averaged rate of apsidal precession and the apsidal
longitudes exhibit phase-locking. The relative apsidal longitudes
is $\Delta\omega=\tilde\omega_{1}-\tilde\omega_{2}$. Malhotra
(2002) described a dynamical mechanism for establishing apsidal
alignment in a pair of planets that initially move on nearly
circular orbits and showed that an impulse perturbation that
imparts an eccentricity to one of the orbits excites the other
planet's eccentricity. Now it is widely believed that the apsidal
resonance plays an important role in the stability of the
discovered multiple-planet systems. In the case of  $\upsilon$
And, the $\Delta\omega$ of the outer two planets of the system
librates about $0^{\circ}$ (Chiang, Tabachnik, \& Tremaine 2001),
also in other systems-GJ 876 (Lee \& Peale 2002) and 47 Uma
(Laughlin, Chambers \& Fischer 2002). The apsidal anti-alignment
of  HD 82943, librating around $180^{\circ}$, was discovered by Ji
\& Kinoshita (in preparation) and  it still occurs for HD 12661
(Lee \& Peale, 2003). However, Beauge et al. (2002) argue that
there may exist the planetary configurations of $\Delta\omega$
librate neither $0^{\circ}$ nor $180^{\circ}$. In the case of  55
Cnc , we found that the $\Delta\omega$ of Companions B and C
librates about $250^{\circ}$ with small amplitude for the
timescale of 3 Myr (see Fig. 2d), and the phenomenon is never
reported before and seems to be a paradigm of the statement by
Beague et al. (2002). In comparison with the outer pairs of
$\upsilon$ And, 55 Cnc differs in that the inner two planets in
the apsidal resonance are simultaneously locked into the state of
the 3:1 MMR. It is likely for the above two mechanisms to be
responsible for the stability of the 55 Cnc system.

Next, what we care is whether the numerical libration of
$\Delta\omega$ of 55 Cnc could be explained in the analytical
viewpoint. Thus we use the L-L secular theory to compute
$\tilde\omega_{1}$ and $\tilde\omega_{2}$ in equation (\ref{eq1})
respectively to show the predicted relative apsidal longitudes. In
Fig. 4,  we note the period of $\Delta\omega$ from theory
$T_{1}=2\pi/(g_{1}-g_{2})\simeq 600$ yr and that of the numerical
integration $T_{2}$ is almost 200 yr. The difference of two
periods again results from the existence of 3:1 MMR. In addition,
it is noteworthy that both of them have the libration width of the
amplitude of $\pm60^{\circ}$. Moreover, by comparing two spans of
libration, we can see the shift of the magnitude of $70^{\circ}$
for the equilibriums, the minimums and maximums of the intervals.
In summary, the analytical curve of the libration confirms the
apsidal resonance found in numerical simulations.

\section{Summary and Discussions}
In this work, we obtain the orbital solutions of the planets of 55
Cnc in the 3:1 resonance. In the simulations, we find that not
only three resonant arguments librate respectively, but
$\Delta\omega$ also librates about $250^{\circ}$. These results
imply the existence of the 3:1 MMR and the  apsidal resonance for
the studied system. We further emphasize that they can act as two
important mechanisms of stabilizing 55 Cnc system. Moreover, it is
the first time to observe the relative apsidal longitude librates
neither $0^{\circ}$ nor $180^{\circ}$. We comment that the
planetary systems in the mean motion resonance always accompany
apsidal alignment, which can play a vital part in maintaining
other non-resonant multiple-planet systems as well.

On the other hand, we argue that the libration of the resonant
arguments may not always occur for even longer timescale, then the
circulation  temporarily takes place for short time and repeat the
progress as time elapses, just as Marcy et al. (2002) pointed out
the 3:1 resonance may appear in the past. However, the progenitor
of the 3:1 MMR for 55 Cnc is not known today. Many works on the
hydrodynamics simulations based on the interaction between the
planets and  disk (Bryden et al. 1999; Kley 2000; Papaloizou,
Nelson\& Masset 2001) are helpful to uncover the mystery.

\acknowledgments {We thank Stan Peale and Xinhao Liao for
insightful discussions. We are grateful to Debra Fischer for
helpful comments to improve the manuscript. This work is
financially supported by the National Natural Science Foundations
of China (Grant 10203005, 10173006, 10233020) and the Foundation
of Minor Planets of Purple Mountain Observatory.}

\clearpage

\begin{deluxetable}{llll}
\tablewidth{0pt} \tablecaption{The orbital parameters of 55 Cnc
planetary system }
%%\tablehead{ \colhead{Parameter} & \colhead{55 Cnc b} &
%%\colhead{55 Cnc c} & \colhead{55 Cnc d} }
\tablehead{ \colhead{Parameter} & \colhead{Companion B} &
\colhead{Companion C} & \colhead{Companion D} } \startdata
$M$sin$i$($M_{Jup}$)\tablenotemark{a}       & 0.83   & 0.20   & 3.69      \\
Orbital period $P$(days)   & 14.65  & 44.27  & 4780      \\
$a$(AU)                    & 0.115  & 0.241  & 5.461     \\
Eccentricity $e$           & 0.03   & 0.41   & 0.28      \\
$\Omega$(deg)              & 138.56 & 306.24 & 227.48    \\
$\omega$(deg)              & 90.25  & 46.39  & 218.91    \\
Mean Anomaly $M$(deg)      & 296.98 & 159.53 & 345.58    \\
\enddata
\tablenotetext{a}{The parameters of the planetary masses, orbital
periods, semi-major axes and eccentricities are taken from Fischer
et al. (2003). The mass of the central star is $1.03M_{\odot}$.}
\end{deluxetable}

\begin{deluxetable}{lrrr}
\tablewidth{0pt} \tablecaption{The values of $e_{ji}$
($j=1,2,3$,respectively for Companions B, C and D; $i=1,2,3$) from
secular Laplace-Lagrange theory for 55 Cnc planetary
system($\times10^8$)} \tablehead{\colhead{$e_{ji}$} &\colhead{1}
&\colhead{2} &\colhead{3\tablenotemark{a}}} \startdata
  $e_{1i}$                  &7641527   &-9022422     &2462\\
  $e_{2i}$                  &5536142   &35711767     &3018      \\
  $e_{3i}$ &   -29    & -18         &27999990  \\
\enddata
\tablenotetext{a}{Note that $e_{13}$ and $e_{23}$ terms are
contributions from Companion D, which is much less than one
percent of the mutual effects of the inner two planets.}
\end{deluxetable}

\clearpage

\epsscale{1.00}
\begin{figure}
\figurenum{1}
\plotone{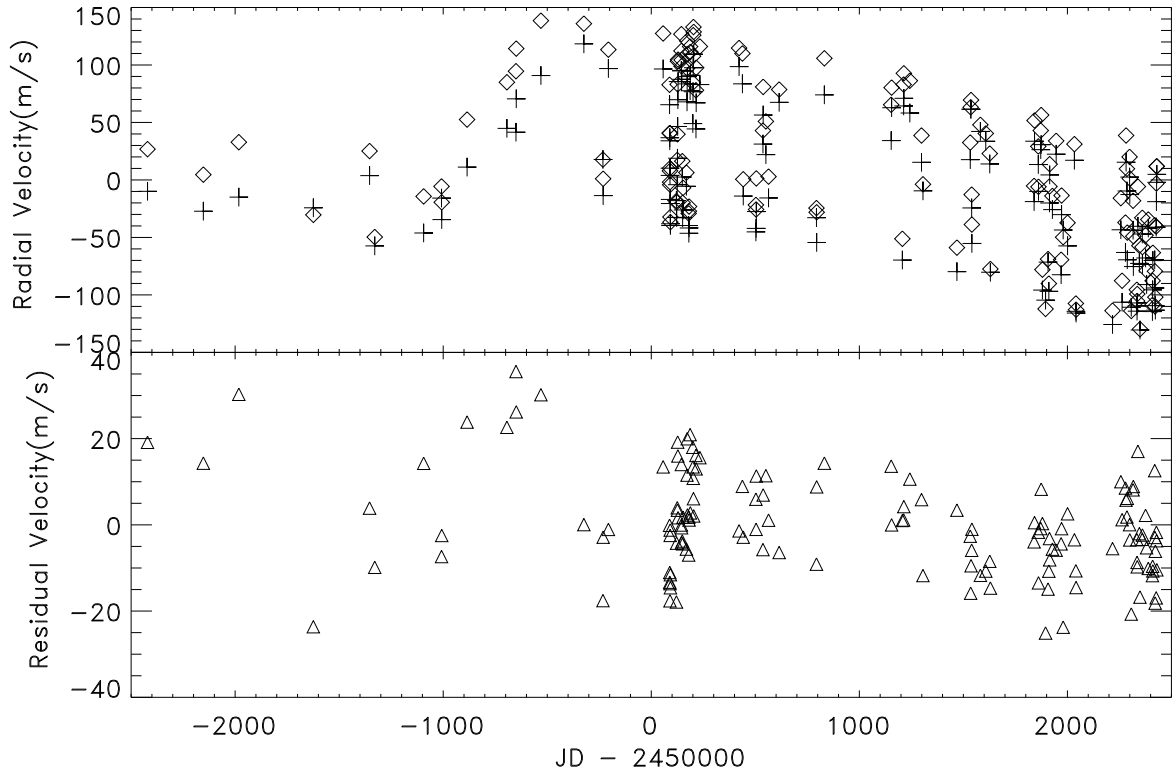} \caption{The upper panel exhibits
the comparison of the measured radial velocities and theoretically
calculated radial velocities for 55 Cnc-the sign diamond
represents the observations given by Marcy et al.(2002), and the
sign plus denotes the results from our orbital solution. The lower
panel exhibits the residual velocity with rms of 11.50 m s$^{-1}$.
\label{fig1} }
\end{figure}
\clearpage

\epsscale{1.00}
\begin{figure}
\figurenum{2}
\plotone{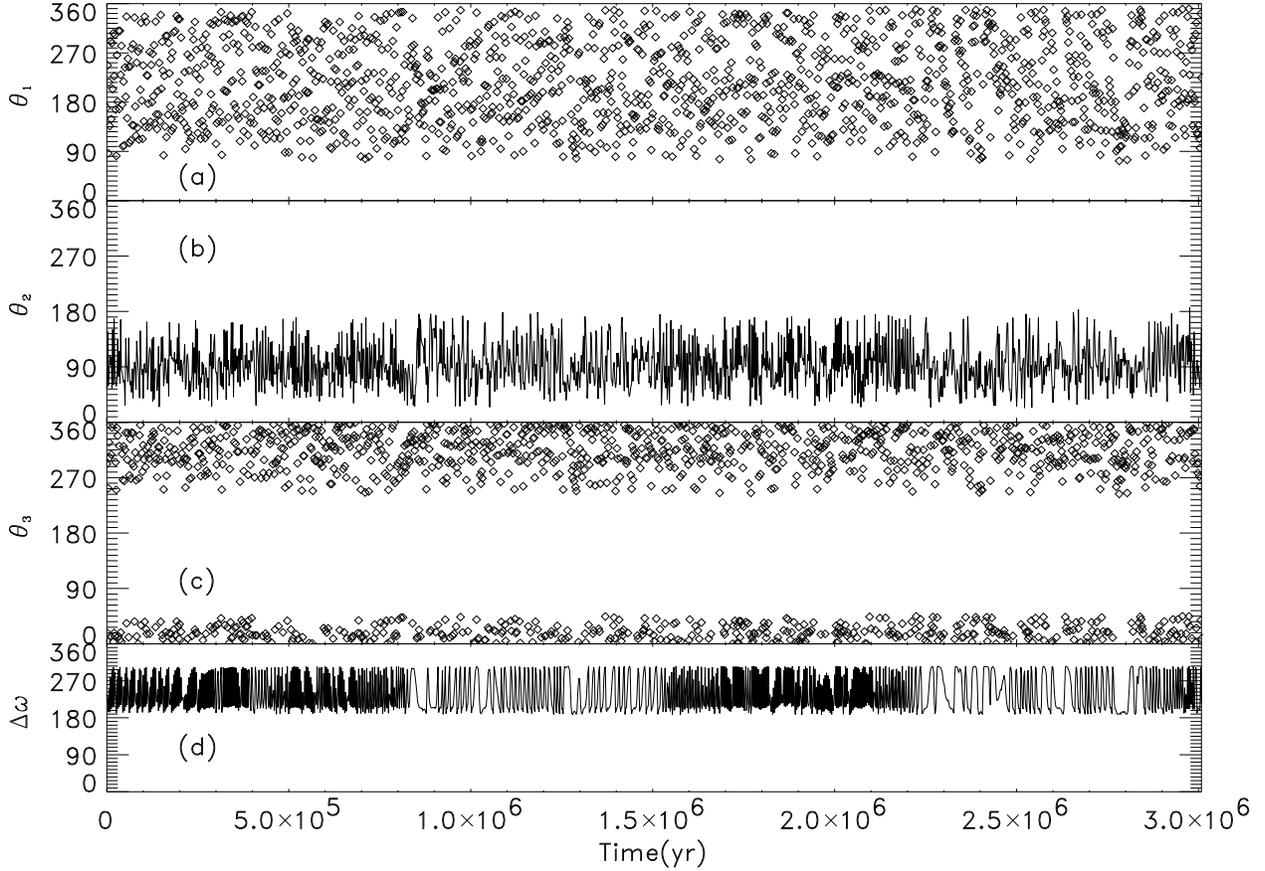} \caption{Panels (a)-(c) are plotted
the variations of $\theta_{1}$, $\theta_{2}$ and $\theta_{3} $
against time. Note that $\theta_{1}$, $\theta_{3} $ respectively
librate about $180^{\circ}$ and $0^{\circ}$ with large amplitudes,
while $\theta_{2}$ librates around $90^{\circ}$ with an amplitude
of $\pm70^{\circ}$ for the timescale of 3 Myr. The libration of
three resonant angles reveals the existence of the 3:1 resonance
of 55 Cnc system. Panel (d) shows the relative apsidal longitudes
$\Delta\omega$ of Companions B and C librates about $250^{\circ}$
with small amplitude for the same timescale. \label{fig2} }
\end{figure}
\clearpage

\epsscale{1.00}
\begin{figure}
\figurenum{3}
\plotone{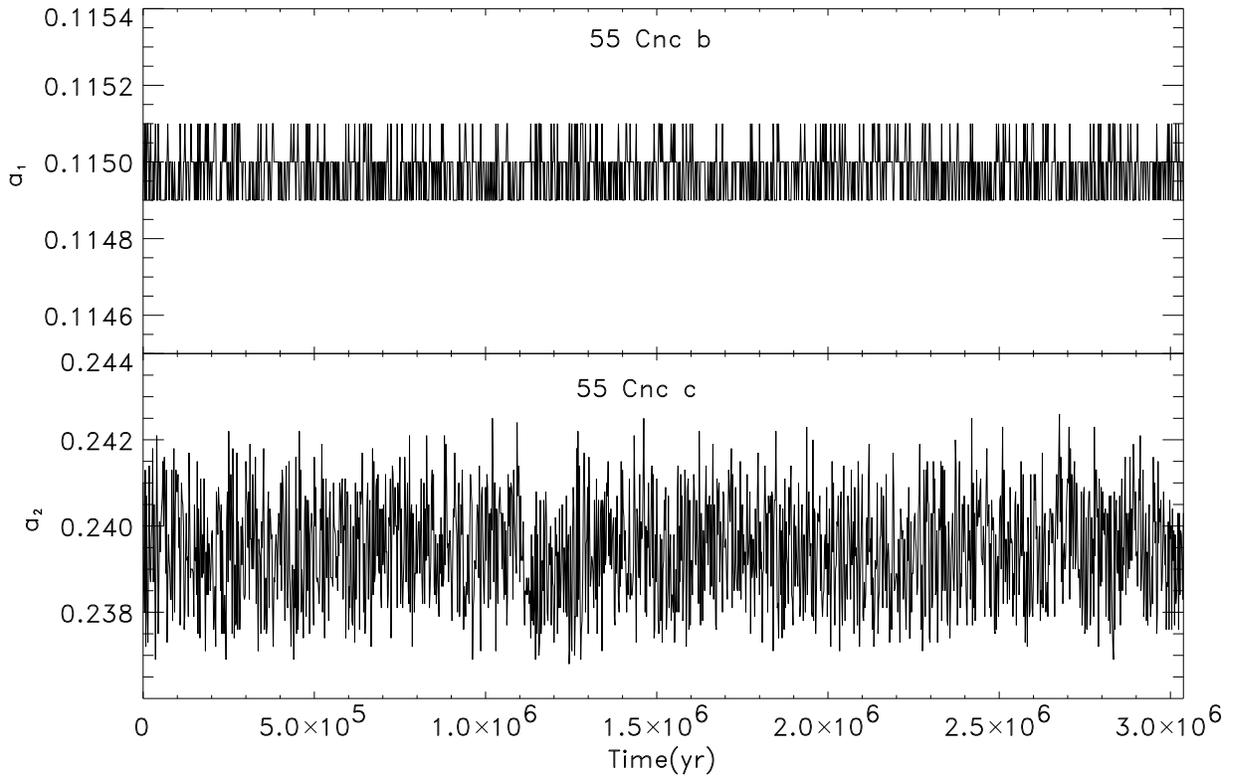} \caption{The upper panel shows that
the semi-major axis $a_{1}$ of Companion B slightly goes around
0.115 AU, while the lower panel displays $a_{2}$ of Companion C
varies about 0.240 AU for millions of years, indicating the two
inner companions are now locked into the 3:1 MMR. \label{fig3} }
\end{figure}

\epsscale{1.00}
\begin{figure}
\figurenum{4} \plotone{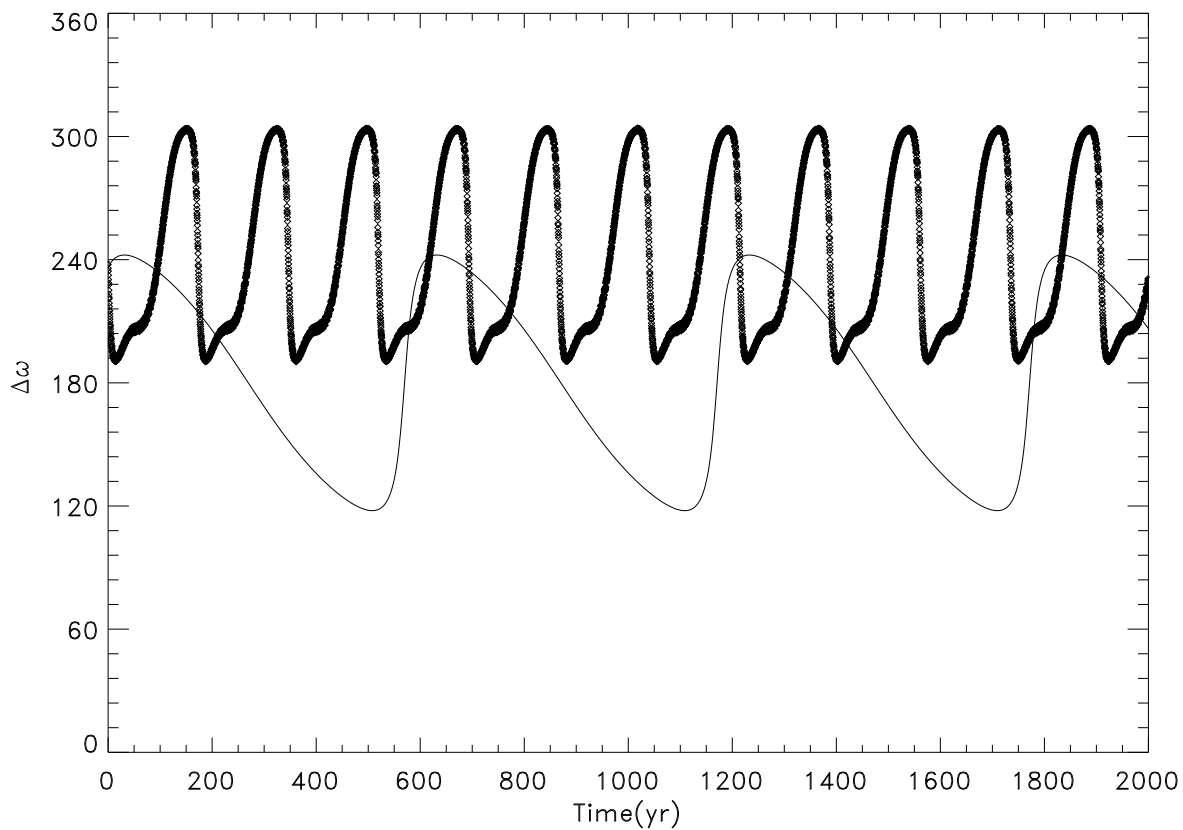} \caption{The libration of the
relative apsidal longitudes of 55 Cnc system. The numerical
results by thick line show $\Delta\omega$ librates about
$250^{\circ}$ degrees with the period of $\sim200$ yr and covers
the span [$190^{\circ}$, $310^{\circ}$]. The thin line represents
the results by Laplace-Lagrange secular theory. See the libration
about $180^{\circ}$ degrees with the period of $\sim600$
yr,locating in the span [$120^{\circ}$, $240^{\circ}$]. Both
exhibit a libration width of the amplitude of $\pm60^{\circ}$.
Notice the shift of the magnitude of $70^{\circ}$ for the
equilibriums, the minimums and maximums of the intervals.
\label{fig4} }
\end{figure}

\clearpage

\end{document}